\documentclass[11pt,twoside]{article}

\usepackage{asp2014}

\aspSuppressVolSlug
\resetcounters

\begin{document}

\title{Improving the visibility and citability of exoplanet research software}

\author{Alice Allen,$^{1,2}$ Alberto Accomazzi,$^3$ and Joe P. Renaud $^{2,4,5}$}
\affil{$^1$Astrophysics Source Code Library (ASCL), Houghton, MI, USA; \email{aallen@ascl.net}}
\affil{$^2$Astronomy Department, University of Maryland, College Park, MD, USA}
\affil{$^3$Center for Astrophysics, Harvard \& Smithsonian, Cambridge, MA, USA}
\affil{$^4$Solar System Exploration Division, NASA/GSFC, Greenbelt, MD}
\affil{$^5$Center for Research and Exploration in Space Science and Technology, NASA/GSFC, Greenbelt, MD}

\paperauthor{Alice Allen}{aallen@ascl.net}{0000-0003-3477-2845}{ASCL/University of Maryland}{Astronomy Department}{College Park}{MD}{20742}{USA}
\paperauthor{Alberto Accomazzi}{aaccomazzi@cfa.harvard.edu}{0000-0002-4110-3511}{Center for Astrophysics, Harvard \& Smithsonian}{Author2 Department}{Cambridge}{MA}{02138}{USA}
\paperauthor{Joe P. Renaud}{joseph.p.renaud@nasa.gov}{0000-0002-8619-8542}{University of Maryland}{Author3 Department}{College Park}{MD}{20742}{USA}



\begin{abstract}
The Astrophysics Source Code Library (ASCL) is a free online registry for source codes of interest to astronomers, astrophysicists, and planetary scientists. It lists, and in some cases houses, software that has been used in research appearing in or submitted to peer-reviewed publications. As of December 2023, it has over 3300 software entries and is indexed by NASA's Astrophysics Data System (ADS) and Clarivate's Web of Science. 

In 2020, NASA created the Exoplanet Modeling and Analysis Center (EMAC). Housed at the Goddard Space Flight Center, EMAC serves, in part, as a catalog and repository for exoplanet research resources. EMAC has 240 entries (as of December 2023), 78\% of which are for downloadable software.

This oral presentation covered the collaborative work the ASCL, EMAC, and ADS are doing to increase the discoverability and citability of EMAC's software entries and to strengthen the ASCL's ability to serve the planetary science community. It also introduced two new projects, Virtual Astronomy Software Talks (VAST) and Exoplanet Virtual Astronomy Software Talks (\emph{exo}VAST), that provide additional opportunities for discoverability of EMAC software resources.
\end{abstract}



\section{Introduction}
In 2020, NASA supported the creation of the Exoplanet Modeling and Analysis Center (EMAC).\footnote{\url{https://emac.gsfc.nasa.gov}} Housed at the Goddard Space Flight Center, EMAC's primary mission is to increase cross-disciplinary collaboration and support scientists working on exoplanet-related projects \citep{EMAC}. EMAC's homepage (see Figure \ref{ex_fig1}) describes the site, shows the number of public resources it contains, and features search and filtering tools to match users with resources that are best suited to their needs. The site currently has 240 entries, 77\% of which are for downloadable software; Figure \ref{ex_fig2} shows an entry for one of the downloadable codes in EMAC.

\articlefigure[width=.90\textwidth]{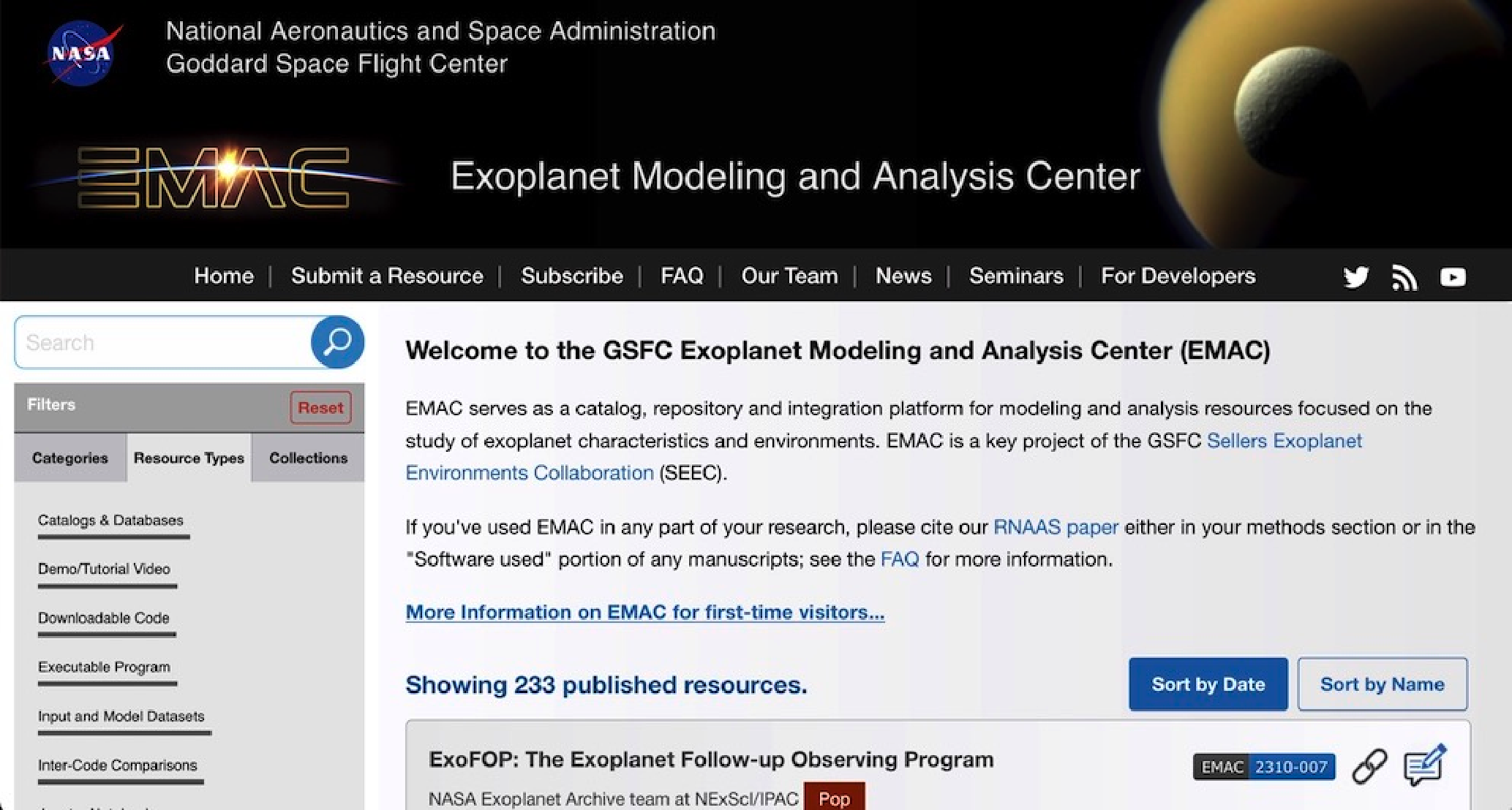}{ex_fig1}{Screenshot of EMAC home page (Nov 2023) showing Resource Types}

\articlefigure[width=.90\textwidth]{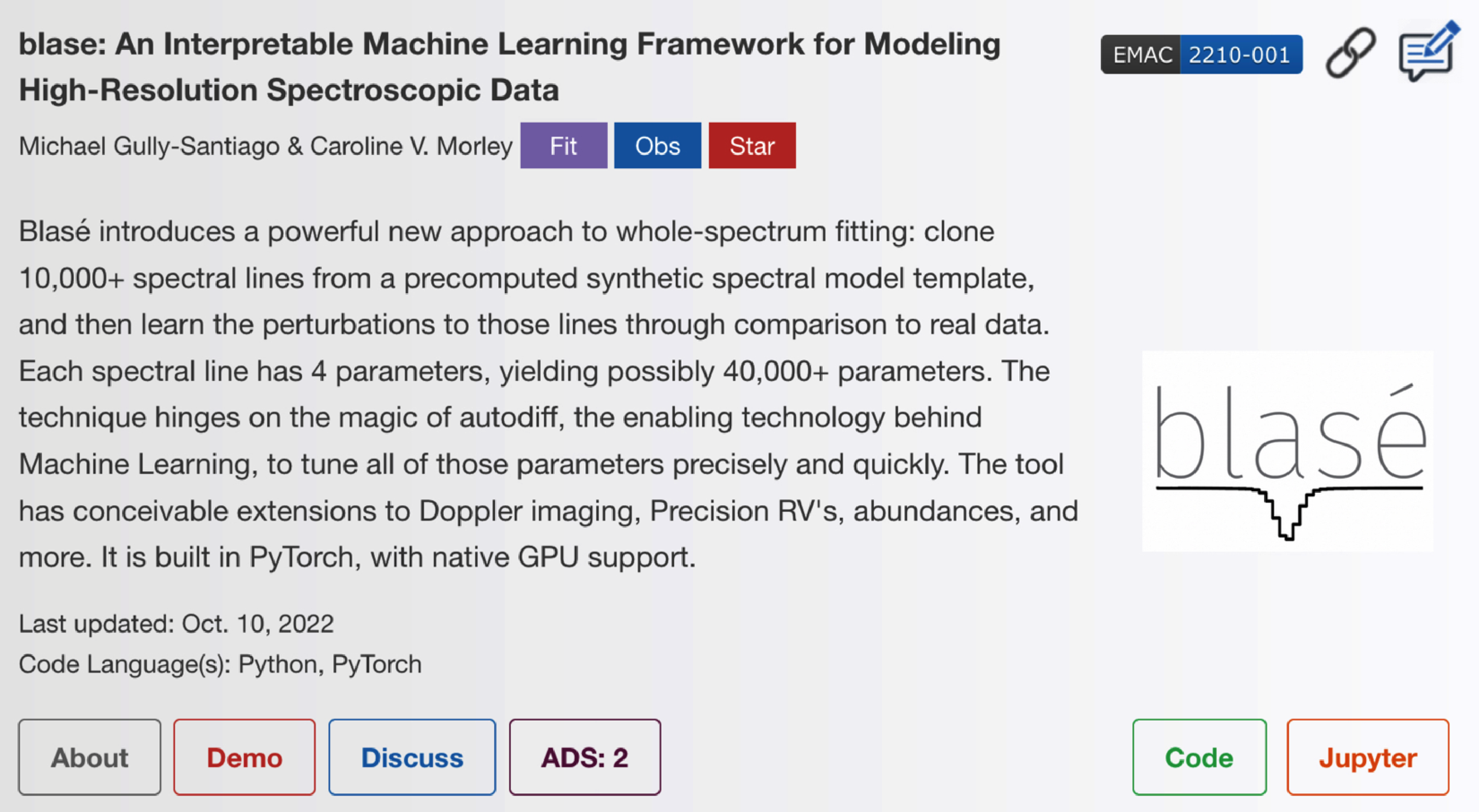}{ex_fig2}{Example of an EMAC downloadable code entry}

\section{Collaborating with others}
In order to raise awareness of the site, EMAC began exploring new collaborations with existing projects, including the ASCL and ADS. 

\subsection{Astrophysics Data System (ADS)}
EMAC personnel approached ADS\footnote{\url{https://ui.adsabs.harvard.edu/}} to ask whether ADS would index EMAC entries. After discussion between the two projects, ADS expressed concern regarding the duplication of records for the same resource, as approximately half of the codes listed in EMAC already have an entry in ADS. Through discussion of different options to achieve the goal of greater visibility for EMAC, ADS came up with a solution that provides visibility without creating duplicate records.

When an EMAC entry for software can be tied to a research article, such as one that describes the code, the ADS button on the EMAC entry (see Figure \ref{ex_fig1}) links to the research paper's entry in ADS. These associations, the links from EMAC to ADS, were leveraged by ADS to add links back to EMAC. As shown in Figure \ref{ex_fig3}, these links appear under ``DATA PRODUCTS'', thus can be used not just for downloadable code entries in EMAC, but also for other resources it holds, such as catalogs.

\articlefigure[width=.85\textwidth]{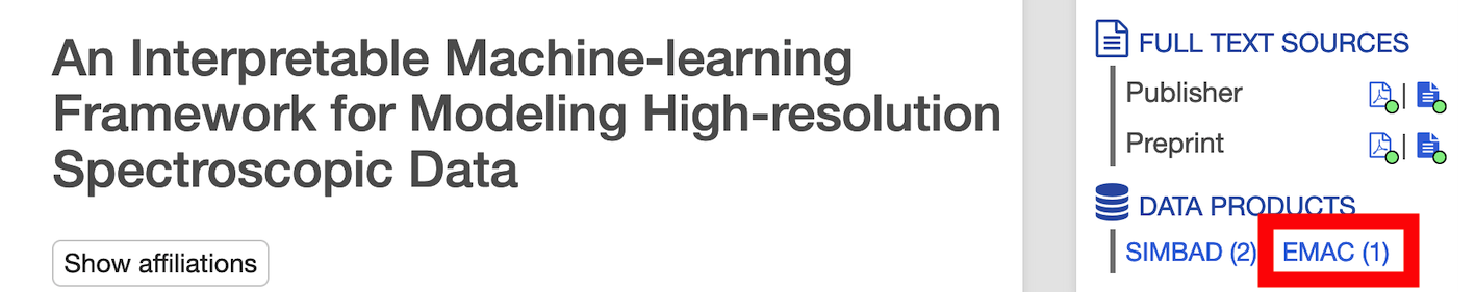}{ex_fig3}{Location of link to EMAC on ADS page}

\subsection{Astrophysics Source Code Library (ASCL)}
One reason approximately half of the downloadable software in EMAC appears in ADS is that the Astrophysics Source Code Library \citep{ASCL_CITE1, ASCL_CITE2} has registered those codes.

The ASCL\footnote{\url{https://ascl.net}} lists and can house research software used in or submitted to peer-reviewed publications. It assigns a unique identifier, the ascl ID, to the codes it registers. The ASCL now has over 3300 software entries. ADS indexes the ASCL, as does the Web of Science.

ASCL entries are citable and citations to them are tracked by indexers; this provides one metric for the impact of research software and accrues credit for these codes to the scientists who write them. ASCL entries are now cited over 2,000 times per year, and over 200 publications indexed by ADS have citations to ASCL records.

ADS and EMAC personnel contacted the ASCL to ask whether ASCL would register downloadable software in EMAC that is not in ASCL. The ASCL is amenable to registering EMAC entries that meet its criteria. Because ADS indexes ASCL, these new entries will be ingested by ADS; their presence in ASCL and ADS improves visibility of these codes and of EMAC. This also helps with citability, as the ASCL provides a unique identifier that can be used to cite the software in a trackable way.

EMAC links its entries to ASCL records, just as it does for ADS; the ASCL would like to change its infrastructure to provide links to EMAC (reciprocal linking), and is seeking funding necessary to do this work.

\subsection{Virtual Astronomy Software Talks (VAST) and the \emph{exo}VAST spin-off}
The Virtual Astronomy Software Talks\footnote{\url{https://vast-seminars.github.io/}} project was started in September, 2022. VAST offers a one-hour seminar on astronomy software on the 3rd Wednesday of each month at 11 AM East coast time (US) via Zoom; Zoom information is sent to the VAST mailing list, which currently has nearly 550 members. The seminar usually features two 20-minute talks followed by discussion, and the seminars are recorded and put on YouTube. Advertisement for the seminar series is typically done through word-of-mouth, email, and social media.

EMAC personnel attended some of the VAST seminars, and at the end of VAST's first year of service, asked whether VAST would support a spin-off just for EMAC-registered software. VAST was amenable to this idea. After an introduction and trial run during a regular VAST seminar, \emph{exo}VAST\footnote{\url{https://vast-seminars.github.io/exovast.html}} started offering seminars in November 2023, also at 11 AM ET (US) via Zoom. It follows the VAST model of hosting two 20-minute talks followed by discussion, with its sessions on the 1st Wednesday of each month, and \emph{exo}VAST also records its seminars and makes them publicly available. 

\section{Current and expected results}
The visibility of EMAC is improved through links from ADS records to EMAC entries and exposure through the VAST and \emph{exo}VAST seminar series. 
Adding suitable EMAC entries to the ASCL provides more places for researchers to come across EMAC resources, not just through use of the ASCL directly but also through the indexing of ASCL entries by ADS, Web of Science, and Google Scholar. 
Further, EMAC software entries become more citable by assigning ASCL IDs to them, enabling their formal citation in research papers. 



\acknowledgements This project uses NASA's Astrophysics Data System Bibliographic Services. The ASCL is grateful for support from the Heidelberg Institute for Theoretical Studies, Michigan Technological University, and the University of Maryland College Park. EMAC support is provided by NASA Goddard Space Flight Center's Sellers Exoplanet Environments Collaboration,\footnote{\url{https://seec.gsfc.nasa.gov/}} a NASA Internal Scientist Funding Model. JPR acknowledges support for this work by NASA under award number 80GSFC21M0002.

\bibliography{C6-06}  


\end{document}